\documentclass[aps,prb,showpacs]{revtex4}

\usepackage{graphicx}
\usepackage{amsmath}

\graphicspath{{./IMAGES/}}

\begin{document}
\narrowtext
\title{Zero-energy states in graphene quantum dots and rings}

\author{C. A. Downing}
\affiliation{School of Physics, University of Exeter, Stocker Road,
Exeter EX4 4QL, United Kingdom}

\author{D. A. Stone}
\affiliation{School of Physics, University of Exeter, Stocker Road,
Exeter EX4 4QL, United Kingdom}

\author{M. E. Portnoi}
\email[]{m.e.portnoi@exeter.ac.uk} \affiliation{School of Physics,
University of Exeter, Stocker Road, Exeter EX4 4QL, United Kingdom}
\affiliation{International Institute of Physics, Av. Odilon Gomes de Lima, 1722, Capim Macio, CEP: 59078-400, Natal - RN, Brazil}

\date{October 28, 2011}

\begin{abstract}
We present exact analytical zero-energy solutions for a class of smooth decaying potentials, showing that the full
confinement of charge carriers in electrostatic potentials in graphene quantum dots and rings is indeed
possible without recourse to magnetic fields. These exact solutions allow us to draw conclusions on the general
requirements for the potential to support fully confined states, including a critical value of the potential strength
and spatial extent.
\end{abstract}


\maketitle
\section{Introduction}

There is a widespread belief that electrostatic confinement of charge carriers in graphene is not possible due to the effect of Klein tunneling,\cite{Klein} by which carriers inside a potential well may couple to states outside the potential via empty states in the hole band. The conventional interpretation of this effect in graphene\cite{Katsnelson,Young} is that the conservation of chirality forbids backscattering for normally incident particles, therefore the transmission probability is unity irrespective of the barrier height. However, the effect is diminished for particles not normally incident on the barrier which can be achieved by particles possessing a finite transversal wave vector.\cite{Silvestrov}

As a result of Klein tunneling, many considerations of axially symmetric potential wells maintain that confinement inside electrostatic quantum dots is impossible.\cite{Chaplik} Instead, focus has been placed on formation of confined states within graphene by the application of magnetic fields perpendicular to the graphene plane\cite{MagMartino,MagGiavaras,MagPeres,MagChen} or on engineering specialized devices which introduce mass-like terms.\cite{Pereira,MassGiavaras,MassNori} Nonetheless, previous authors have studied the manipulation of charge carriers by electrostatic fields such as the creation of quasibound states in quantum dots, whose lifetimes are long but not infinite, as certain angular momenta correspond to trajectories incident on the potential barrier with low transmission probabilities.\cite{MagChen,Matulis} An alternative interpretation of long-lived states in electrostatic dots is that a well-tuned structure can generate wavefunctions outside the dot which interfere destructively.\cite{Hewageegana}

It is well known that an ideal two-dimension system with linear energy dispersion possesses a density of states that vanishes in the limit of zero energy.\cite{CastroNeto} This raises the possibility of confining states by careful manipulation of the Fermi energy by application of a back-gate voltage, allowing the energy of the system to coincide with the Dirac points. In this scenario, no states exist outside of an electrostatic potential well for bound states to tunnel into. This approach has previously been developed with reference to electron waveguides,\cite{SmoothWaveguides} for tunneling selection of charge carriers through an n-p junction,\cite{Cheianov} and for sharply terminated wells of constant depth.\cite{Titov}
In this paper we look at full confinement via this method in smooth potentials with circular symmetry. We would like to emphasize that our analytic results are obtained within an idealized model neglecting impurity- and substrate-induced disorder, local strain distribution, ripples, and trigonal warping as well as many-body and finite-size effects, which all influence the energy spectrum and transport properties of realistic graphene flakes, especially near the charge neutrality point.\cite{CastroNeto}

The application of an external field with circular symmetry would be possible by close proximity of a scanning tunneling microscope (STM) tip. The presence of bound modes can then be probed by performing conductivity measurements, in which electrons may propagate between terminals via the states created by the potential. Similar experiments relevant to the creation of localized states have recently been performed.\cite{Ensslin,Berezovsky}

Quantum dots are important for the understanding of fundamental physics and also for applications, for example, being crucial to progress in spintronics and quantum computation. Applications of graphene to these disciplines is doubly important: first because graphene is considered to be a system where  zero-energy Majorana fermions may appear,\cite{Majorana} and second because quantum dots in graphene have potential for use as spin qubits.\cite{Recher}
Notably, most of the results of this work are also applicable to the Dirac-like states on the surface of topological insulators,\cite{KaneRMP10} which is another exciting area of contemporary condensed matter physics.

The Hamiltonian operator in the massless Dirac-Weyl model for graphene, which describes the motion of a single electron in the presence of an external axially symmetric potential $U(r)$, is \cite{McClure,Novoselov05}
\begin{equation}
\label{DiracEq}
\hat H = v_{\mathrm F} \boldsymbol \sigma \cdot \boldsymbol{\hat p} + U(r),
\end{equation}
where $v_{\mathrm F}\approx c/300$ is the Fermi velocity of the charge carriers, $\boldsymbol \sigma=(\sigma_x,\sigma_y)$ are the Pauli spin matrices, and $\boldsymbol{\hat p}=-i\hbar\nabla$ are the linear momentum operators. We consider smooth confining potentials, which do not mix states in the two nonequivalent graphene valleys, and deal with one valley only. All our results can be easily reproduced for the other valley. For axial symmetry, we transform into cylindrical coordinates $(r, \theta)$ and separate the variables by using the following ansatz for the two-component wavefunction
\begin{equation}
\label{CompleteWF}
\Psi(r,\theta) = \frac{e^{im\theta}}{\sqrt{2\pi}} \left(
 \begin{array}{c}
 \chi_A(r) \\ e^{i\theta}\chi_B(r)
 \end{array}
\right),
\end{equation}
where $m$ is the integer-valued angular momentum number and the subscripts A and B refer to the sublattices of the graphene honeycomb lattice. This choice of wavefunction leads to two coupled first order differential equations for the radial wavefunction components $\chi_{A,B}$
\begin{align}
\label{coupledchi}
 \begin{split}
  \left(-\frac{d}{dr} - \frac{m+1}{r} \right) i\chi_B &= (\varepsilon-V(r))\chi_A, \\
  \left(-\frac{d}{dr} + \frac{m}{r}   \right) i\chi_A &= (\varepsilon-V(r))\chi_B,
 \end{split}
\end{align}
with $V(r)=U(r)/\hbar v_{\mathrm{F}}$ and $\varepsilon=E/\hbar v_{\mathrm{F}}$, where $E$ is the energy eigenvalue. It is most convenient to solve these equations by formulating a second order equation for one component and reusing one of Eqs.~\eqref{coupledchi} to find the other wavefunction component. It is worth noting that the radial wavefunction components transform into each other on replacing $m\to -(m+1)$, which is important when considering special values of $m$.

The rest of the paper is organized as follows. In Sec.~\ref{LongRange}, we look into the long-range behavior of zero-energy states in power-law potentials decaying at large distances faster than the Coulomb potential and show that only wavefunctions corresponding to a non-zero angular momentum (vortices) are square-integrable across the whole graphene plane. Section~\ref{Exact} contains the main results of our work -- hitherto unknown analytic solutions for fully confined zero-energy states in a class of circularly symmetric decaying potentials. In Sec.~\ref{STM} we discuss the relevance of our results to STM of graphene. Finally, in Sec.~\ref{Disc} we summarize our results and formulate possible further applications and developments of our work.

\section{Long-Range Behavior in a Decaying Potential}
\label{LongRange}
The Coulomb problem for Dirac fermions (the Dirac-Kepler problem) in two dimensions has been studied for some time,\cite{Yang} and there has been a resurgence of interest in this problem since the discovery of graphene.\cite{ShytovVacpol, Nilsson, Novikov} Despite its apparent beauty, the ideal Coulomb problem has somewhat questionable relevance to the reality of graphene since the impurity potential should have a short-range cutoff of the Ohno type\cite{OhnoExcitons} as well as a faster than $1/r$ long-range decay due to either screening or an image charge appearing in the metallic back gate, which is necessarily present in graphene-based devices.

For a more realistic quickly decaying potential when a system is fixed by the back-gate at the Dirac point energy ($E=0$), there is no coupling to the continuum and, as we show, a single fully bound state might appear. General properties of the fully confined solutions can be understood from their long-range behavior. Thus we consider the general potential given by
\begin{equation}
\label{powerpotential}
V(r)=V_0 r^{-p},
\end{equation}
where the rate at which the potential falls off is characterized by $p>1$. For this potential the differential equation obeyed by the upper wavefunction component is
\begin{equation}
\label{powerODE}
\chi_A'' + \left(\frac {1+p}{r} \right) \chi_A' + \left( \frac{V_0^2}{r^{2p}} - \frac{m(p+m)}{r^2} \right) \chi_A = 0.
\end{equation}

This equation can be recast as the Bessel differential equation. In general the solution is a linear combination of Bessel functions in the form
\begin{equation}
\label{BesselSolA}
\chi_A(r) = r^{-p/2} \left[ c_1 J_\alpha \left( \frac{V_0}{p-1}r^{1-p} \right) + c_2 J_{-\alpha} \left( \frac{V_0}{p-1}r^{1-p} \right) \right],
\end{equation}
where $\alpha=(p+2m)/(2-2p)$ is the order of the function. Then it follows from Eqs.~\eqref{coupledchi} that the lower radial wavefunction component is
\begin{equation}
\label{BesselSolB}
\chi_B(r) = i r^{-p/2} \left[ c_1 J_{\alpha+1} \left( \frac{V_0}{p-1}r^{1-p} \right) - c_2 J_{-\alpha-1} \left( \frac{V_0}{p-1}r^{1-p} \right) \right].
\end{equation}

Now we look into the long-distance behavior of these solutions in order to gain insight into the behavior of carriers in a general confining potential. For large radial coordinate, the variable of the Bessel functions tends to 0, thus the desired behavior is dominated by the first-order term in the Maclaurin expansion. For $m\geq0$ the asymptotic solution becomes
\begin{align}
\label{Asymppos}
\chi_A(r) \sim \frac {c_2} {r^{m+p}}, \qquad
\chi_B(r) \sim \frac {ic_2}{r^{m+1}}.
\end{align}
Similarly for $m\leq-1$ the asymptotic solution is
\begin{align}
\label{Asympneg}
\chi_A(r) \sim c_1 r^m, \qquad
\chi_B(r) \sim ic_1 r^{1+m-p}.
\end{align}

From this asymptotic behavior of the wavefunctions we can deduce that all states in any smoothly decaying potential are normalizable, with the exception of angular momentum states $m=0$ and $m=-1$, reflecting the $m\to -(m+1)$ symmetry. In these cases the integral of the probability distribution contains a logarithmically divergent term. All other angular momenta describe quasiclassical trajectories for which charge carriers are not normally incident on the confining potential (vortices), consistent with the idea that Klein tunneling in graphene is maximized at normal incidence and suppressed at other angles.\cite{Katsnelson,Titov,Matulis}

It is worth noting that for $p$ equal to an even integer, for at least a subset of $m$, the two solutions given in Eq.~\eqref{BesselSolA} are not linearly independent and the second independent solution for the wavefunction component is instead a Neumann function with the same order ($\alpha$) as the first. However the asymptotic behavior of this function is the same as for the Bessel function with order $-\alpha$, so the conclusions we draw are unchanged.

\section{Smooth Quantum Dots and Rings}
\label{Exact}
We are interested in the presence of confined states when the system is fixed at the Dirac point energy, consistent with $\varepsilon=0$ in the low-energy formulation. We consider the class of smoothly-varying radial potentials
\begin{equation}
\label{lorentzrings}
V(r) = \frac{ V_0 (r/d)^k }{ 1 + (r/d)^{2(k+1)} }
\end{equation}
where $V_0$ and $d$ parametrize the strength and width of the potential respectively and $k$ determines the sharpness of the ring. Most notable is the case where $k=0$, corresponding to a potential energy profile of the well-known Lorentzian shape. The benefit of a potential in this form is that it contains parameters that allow it to be fitted to experimentally realizable potentials and that it is both regular at the origin and short-range. The smoothness of the potential also allows one to neglect the intervalley scattering that occurs for tunneling problems, contrary to previous considerations of quantum dots with sharp boundaries.\cite{Chaplik,Titov,Matulis}

For simplicity, we first consider the case of the quantum dot, $k=0$. For the upper component of the wavefunction we seek a solution in the form
\begin{equation}
\label{ansatz}
\chi_A(r) = \left(\frac{r}{d}\right)^m \eta(r),
\end{equation}
which leads to the second-order differential equation for $\eta(r)$,
\begin{equation}
\label{etarODE}
\eta''(r) + \left( \frac{2m+1}{r} + \frac{2r/d^2}{1+(r/d)^2} \right)\eta'(r) + \left(\frac{V_0}{1+(r/d)^2}\right)^2\eta(r) =0,
\end{equation}
and changing to a new variable $z=\left(r/d\right)^2$ we obtain an equation for $\eta(z)$,
\begin{equation}
\label{etaODE}
z(1+z)^2\eta''(z) + (1+z) \big[ (m+1)(1+z)+z \big]\eta'(z) + \left( \frac{V_0d}{2} \right)^2 \eta(z) =0.
\end{equation}

This differential equation has the following solution for $m\geq0$, which can be written using appropriate 
identities,\cite{AbramAndSteg} as
\begin{equation}
\label{etahyper1}
\eta(z) = c_1 ~_2F_1\left(n,-n;m+1;\frac{1}{1+z^{-1}}\right),
\end{equation}
where $n=V_0d/2$ must take a non-negative integer value in order to terminate the hypergeometric function and $c_1$ is a normalization factor. We can see that the solution is unchanged upon replacing $V_0$ with $-V_0$, as required for electron-hole symmetry. In order to have physically meaningful solutions we require that $n>m$, ensuring that the radial wavefunctions do not diverge.

For $m\leq-1$ the solution is
\begin{equation}
\label{etahyper2}
\eta(z) = c_2 ~_2F_1\left(n,-n;-m;\frac{1}{1+z}\right),
\end{equation}
for which the corresponding requirement of the presence of physical wavefunctions is $n\geq-m$.

The states formed by the presence of a Lorentzian-type electrostatic potential acting on a graphene sheet do indeed fulfill the expectation presented in Sec.~\ref{LongRange}, that the states with $m=0$ and $m=-1$ are extended states with a nonintegrable probability density. Thus, the first normalizable state occurs for $m=1$ (or $m=-2$), which requires $n=2$ and so the threshold for fully confined states is $V_0d=4$. When the potential is fixed, charge carriers fill up modes enumerated by $m$. The apparent asymmetry between positive and negative values of $m$ is removed when the other graphene valley is considered.

A similar procedure allows one to solve the coupled equations with the potential describing electrostatic rings, namely, the potential \eqref{lorentzrings} with $k \neq 0$. The general solution in this case is found to be
\begin{align}
\label{etahypergeneralk}
\begin{split}
 \eta(r) &= c_1 ~_2F_1\left(p_1,-p_1; q_1; \frac{1}{1+\xi^{-1}} \right) 
 + c_2 ~_2F_1\left(p_2,-p_2; q_2; \frac{1}{1+\xi} \right),\qquad \\
 q_1 &= \frac{k+2m+2}{2k+2}, \qquad
 q_2 = \frac{k-2m}{2k+2},   \qquad
 \xi = \left(\frac{r}{d}\right)^{2k+2}.
\end{split}
\end{align}

These wavefunctions strongly resemble the $k=0$ case. For $m\geq0$ the second term is unphysical and requires one to set $c_2=0$; likewise for $m\leq-1$ one requires that $c_1=0$. The termination of the hypergeometric series requires
\begin{equation}
p_{1,2}=\frac{V_0d}{2k+2} = N + q_{1,2},
\end{equation}
where $N$ is a non-negative integer which corresponds to the number of non-zero nodes in the wavefunction. There are no additional requirements for the regularity of the wavefunction.
\begin{figure}[h]
\includegraphics[width=0.35\textwidth]{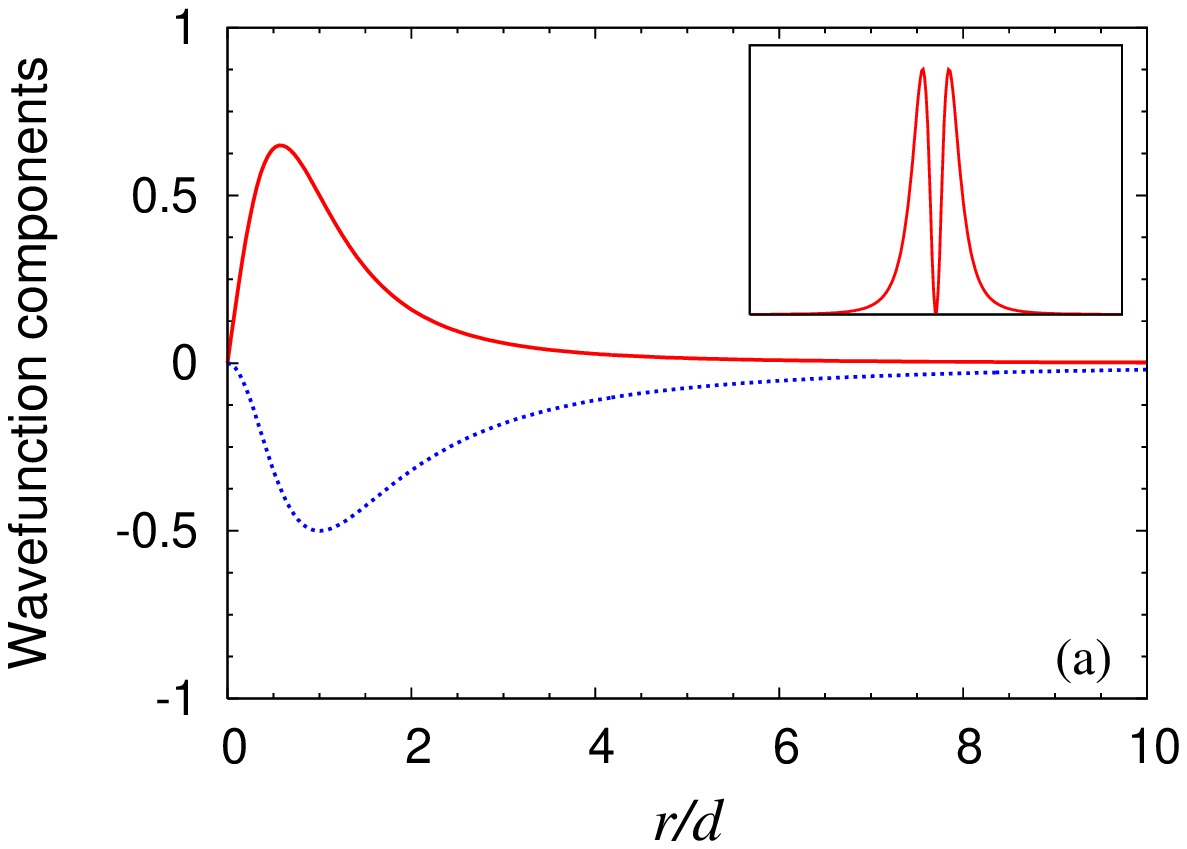}
 \includegraphics[width=0.35\textwidth]{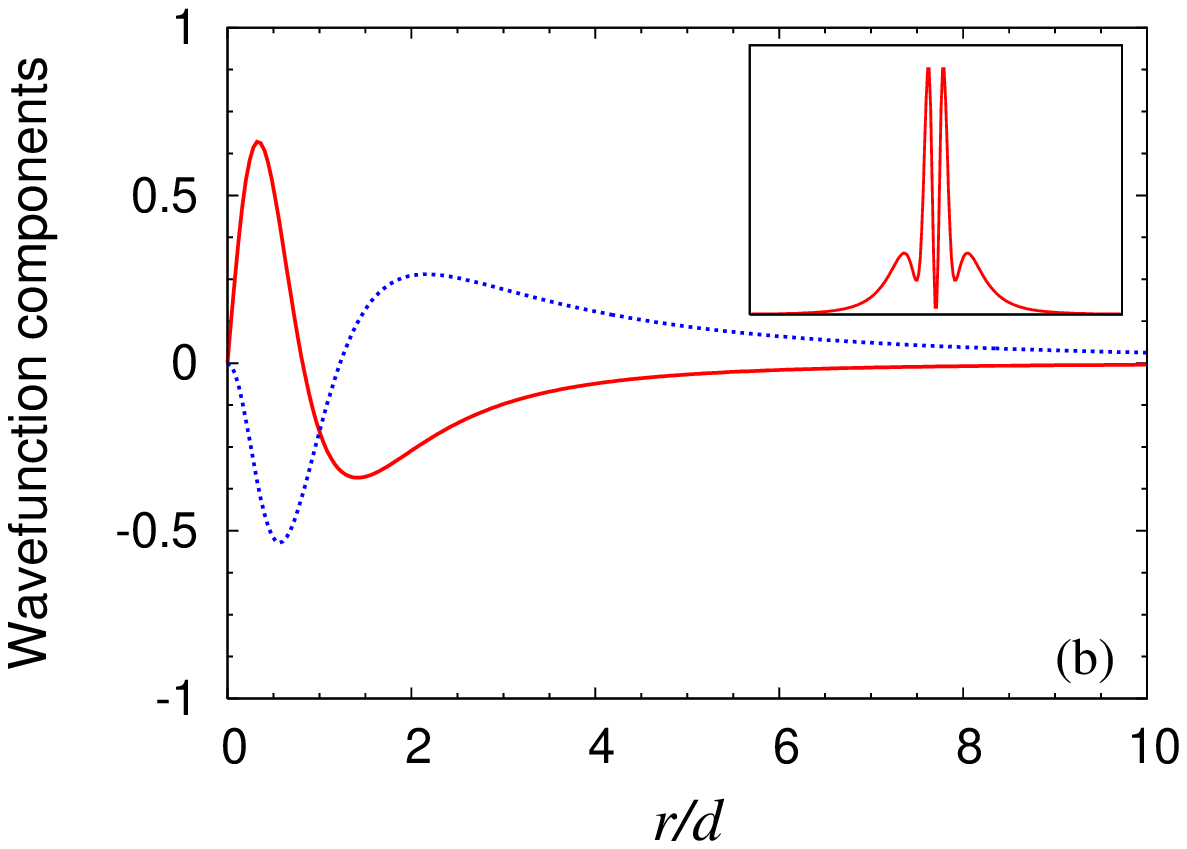}
 \includegraphics[width=0.35\textwidth]{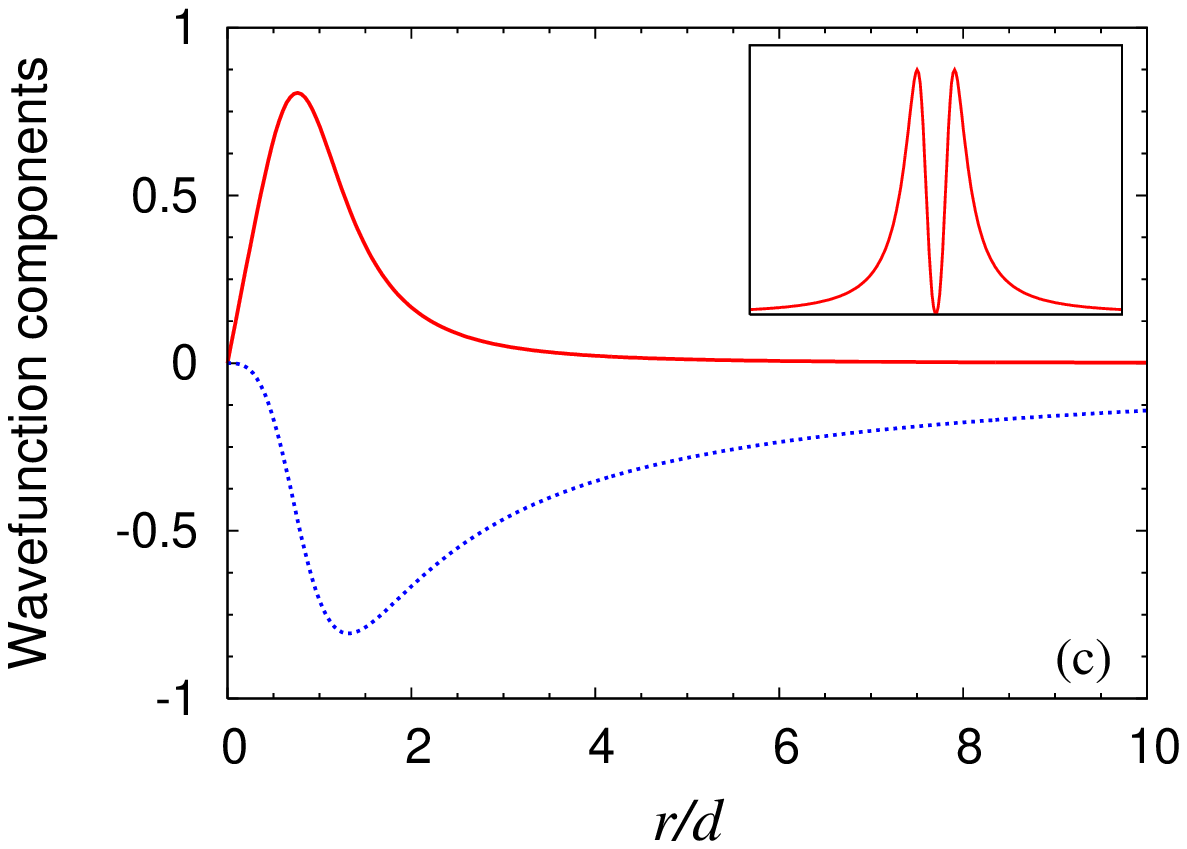}
 \includegraphics[width=0.35\textwidth]{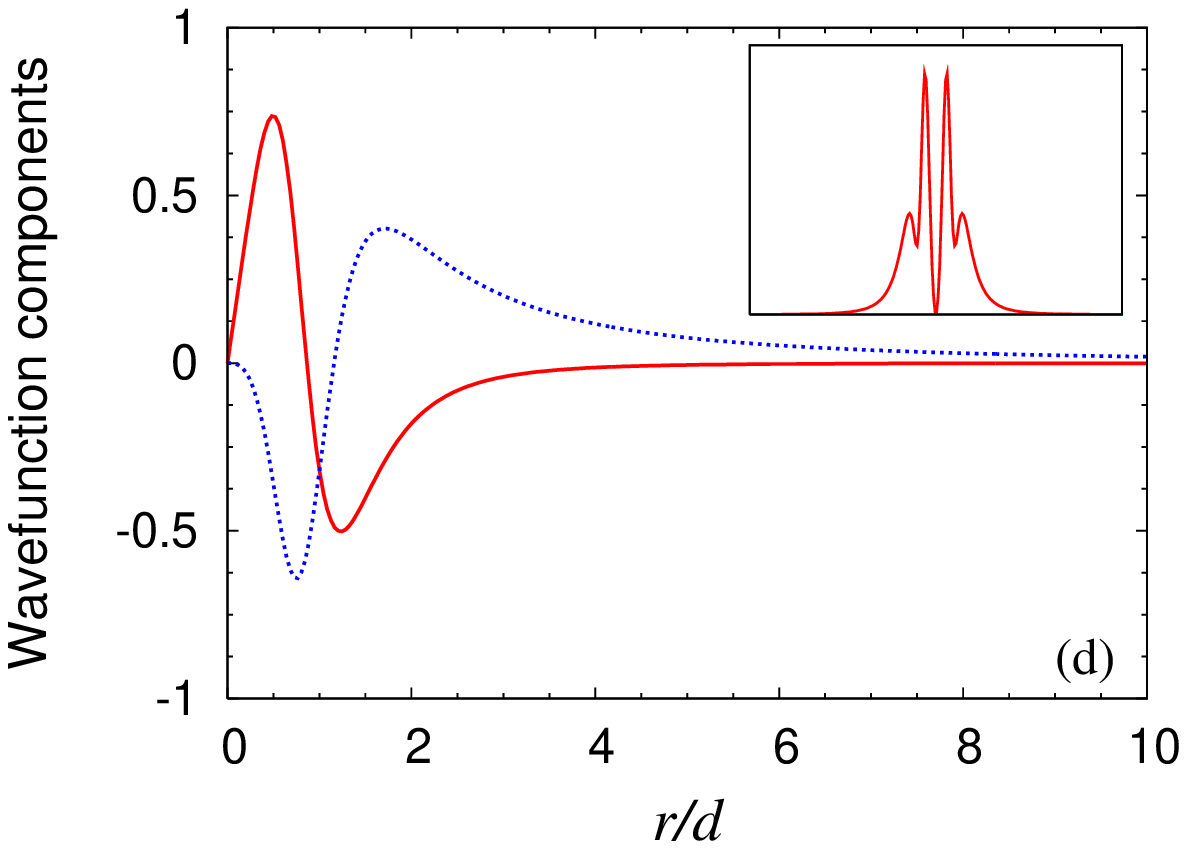}
 \caption{(Color online) Radial wavefunction components for the first two states ($N=0,1$) with angular momentum $m=1$ for the Lorentzian dot ($k=0$) and the first ring ($k=1$): (a) $k=0$, $N=0$; (b) $k=0$, $N=1$; (c) $k=1$, $N=0$; and (d) $k=1$, $N=1$. Solid (dotted) lines correspond to components $\chi_A$ ($i\chi_B$). Insets: shape of the probability density for each state.}
 \label{WFs}
\end{figure}

The radial wave-function components for $m=1$, the first angular momentum to permit confined modes, are plotted in Fig.~\ref{WFs} for both the Lorentzian dot and the simplest $k=1$ ring. As an example the first fully integrable mode, which corresponds to node number $N=0$ with $m=1$ for the Lorentzian dot, has the following normalized two-component wavefunction
%\begin{equation}
%\label{examplewf}
%\Psi(r,\theta) =\frac{1}{d}\sqrt\frac{2}{\pi}\frac{\frac{r}{d} e^{i\theta}}{\left(1+\frac{r^2}{d^2}\right)^2} \left(
% \begin{array}{c}
%  1 \\
%  -i \frac{r}{d} e^{i\theta}
%\end{array}
%\right).
%\end{equation}
\begin{equation}
\label{examplewf}
\Psi(r,\theta) =\frac{1}{d}\sqrt\frac{2}{\pi}\frac{\left(r/d\right) e^{i\theta}}{\left[ 1+(r/d)^2\right]^2} \left(
 \begin{array}{c}
  1 \\
  -i \frac{r}{d} e^{i\theta}
 \end{array}
\right).
\end{equation}

\section{Probe Microscopy}
\label{STM}
\begin{figure}[h]
 \includegraphics[width=0.60\textwidth]{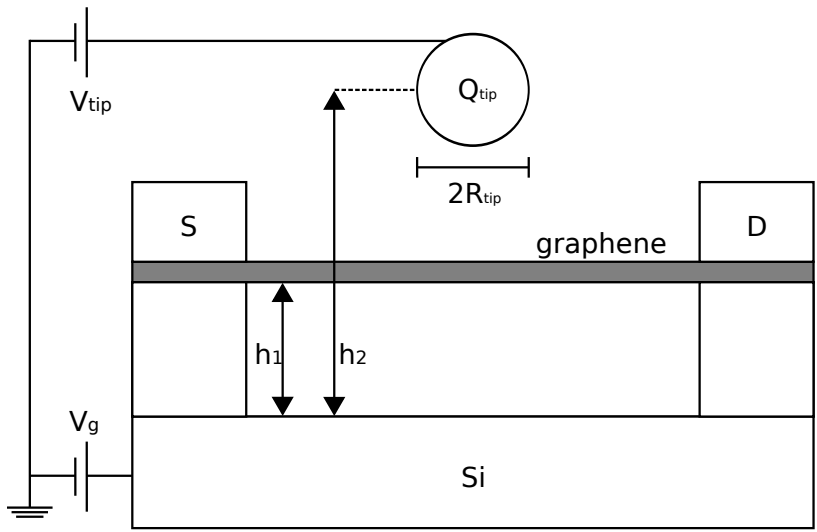}
 \caption{Schematic representation of a proposed experimental setup in which a charged STM tip with characteristic radius $R_\mathrm{tip}$ produces confined states in the graphene sheet either by varying the charge on the tip or by fixing the tip voltage $V_\mathrm{tip}$ and moving the tip vertically with respect to the flake. The presence of bound states is exhibited by peaks in the conductivity whenever new states appear.}
 \label{STMsetup}
\end{figure}

The immediate physical relevance of the electrostatic Lorentzian potential becomes apparent by considering a standard image charge problem.\cite{GriffithsEM} We construct a toy model of an STM tip in which the tip is treated as a spherical charge held above the graphene plane at the charge neutrality point (i.e., $E=0$), depicted schematically in Fig.~\ref{STMsetup}, which produces the potential
\begin{equation}
\label{toySTM}
U_{\mathrm{STM}}(r) \approx \frac{e Q_\mathrm{tip}}{4 \pi \kappa} \left( \frac {1}{ \sqrt{r^2+(h_1-h_2)^2} } - \frac {1}{ \sqrt{r^2+(h_1+h_2)^2} } \right),
\end{equation}
where $\kappa$ is the permittivity of the air, $Q_\mathrm{tip}$ is the total charge accumulated on the STM tip and $h_1$ and $h_2$ are the distances from the metallic surface to the graphene layer and STM tip, respectively. Clearly the electrical potential energy experienced by charge carriers in the graphene plane is independent of the radius of the spherical tip when the charge is fixed, as follows from Gauss' law. The charging of the tip by an applied voltage is, however, dependent on the specific shape of the system via the capacitive coupling between the tip and the Si layer.

Evidently, this form of the potential created by the tip only includes the leading-order term due to point charges. It is relatively simple to include higher order terms that obey the boundary condition of a fixed potential on the surface of the tip by positioning an infinite number of imaginary point charges inside the charged tip and its image.\cite{Brazilians} However these higher order terms do not significantly affect the results.

\begin{figure}[tbp]
 \includegraphics[width=0.6\textwidth]{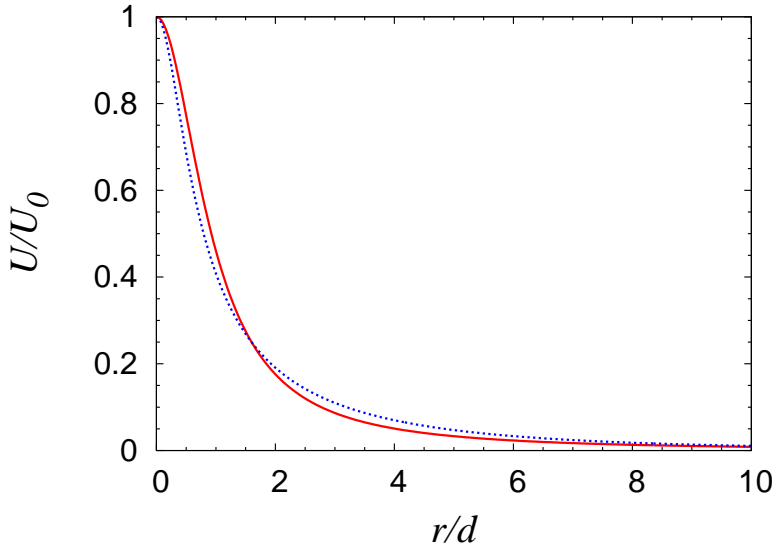}
 \caption{(Color online) Comparison of a repulsive electric potential created by a charged STM tip (blue dotted line) to the solvable Lorentzian potential (red solid line) as a function of scaled radial distance. The potential profiles clearly are very similar. Chosen values of $h_1$ and $h_2$ here are arbitrary; the profiles resemble that shown here for any values so long as $h_2>h_1$. $U_0=V_0\hbar v_\mathrm{F}$ is the maximum value of the potential at the origin.}
 \label{PotCompare}
\end{figure}

The Lorentzian dot potential can trivially be fitted to Eq.~\eqref{toySTM} by equating the potential maxima at the origin and the integrals of the two potentials, yielding the matching criteria,
\begin{equation}
\label{matching}
 V_0= \frac{e Q_\mathrm{tip}}{4 \pi \kappa \hbar v_\mathrm{F}} \frac{2h_1}{h_2^2-h_1^2}
\end{equation}
and
\begin{equation}
\label{matching2}
   d= \frac{h_2^2-h_1^2}{\pi h_1} \ln\left( \frac{h_2+h_1}{h_2-h_1} \right).
\end{equation}
The result of the fitting procedure using Eqs.~(\ref{matching}) and (\ref{matching2}) is shown in Fig.~\ref{PotCompare}. It is experimentally feasible to achieve a probe tip with a radius of curvature $R_\mathrm{tip} \approx 20\mathrm{nm}$ and to set the height of the tip above the graphene plane of the order of $10\mathrm{nm}$ without deforming or damaging the graphene sheet.\cite{Berezovsky} The thickness of the air gap is of the order of hundreds of nanometers, such that the capacitance between the tip and its image is almost independent of the separation distance between the tip and the graphene flake, and so it can be approximated to
\begin{equation}
\label{capacitance}
C_\mathrm{tip} \approx 4\pi\kappa R_\mathrm{tip}.
\end{equation}
Thus we expect the values of the tip voltage at which confined states occur to be given, to a good approximation, by
\begin{equation}
\label{prediction}
V_\mathrm{tip} \approx \frac{ n\pi \hbar v_{F} }{e R_\mathrm{tip}\ln\left( \frac{h_2+h_1}{h_2-h_1} \right)},
\end{equation}
where $V_\mathrm{tip}=Q_\mathrm{tip}/C_\mathrm{tip}$. As long as the tip does not come too close to the graphene plane, the voltage needed to form states is only weakly affected by the distance from the graphene, and instead the radius of the tip is the dominant factor. This simplified model suggests a tip of radius $20\,\mathrm{nm}$ held at height $20\,\mathrm{nm}$ above the graphene plane, suspended over an air gap of $280\mathrm{nm}$ (dimensions that are experimentally feasible)\cite{Berezovsky} would create confined states at integer multiples of $38\mathrm{mV}$. According to Ref.~[\onlinecite{Titov}], the emergence of confined states will be accompanied by conductivity peaks in the sample maintained at zero Fermi energy by the back gate.

\section{Discussion}
\label{Disc}
We have shown, using a model class of potentials, that full confinement of the charge carriers in graphene by electrostatic potentials is indeed possible for zero-energy states, so that the Fermi level coincides with the Dirac points. This setup can circumvent the effect of Klein tunnelling providing a new way to create quantum dots. The exact solutions shown here, describing both model quantum dots and rings, allow one to understand more general properties of such systems. Namely, square-integrable states are only possible when the angular momentum quantum number $m$ is non-zero, and, due to the sublattice symmetry, also when $m\neq-1$. Thus the fully confined states are zero-energy vortices, which require for their existence the critical effective strength of the confining potential (the product of the potential strength and characteristic size). We have analyzed the feasibility of experimentally probing the presence of confined states in quantum dots induced by an STM tip above the graphene surface.

The implications of our results go beyond the discussed STM geometry and can be used for a better understanding of the zero-energy states in undoped graphene. Indeed a circularly symmetric potential induced by impurities should also support zero-energy vortices upon certain conditions. At first glance full confinement can only be achieved by fine-tuning the potential depth and width. However the natural tendency of the system towards local neutrality, which minimizes the total energy including the energy associated with the electrostatic field, allows one to speculate that the screening would adjust the potential parameters in favor of the existence of zero-energy states. Notably, this will create non-zero charge density at the charge neutrality point in disordered samples providing another plausible cause of the minimal conductivity of graphene.\cite{CastroNeto} The discussed non-linear screening should also ease the observation of fully-confined states in STM spectroscopy, as it will adjust the strength and shape of the potential to support zero-energy states for an interval of tip voltages instead of a fixed voltage value.

Our results also provide some insight into the problem of zero-energy excitons in pristine graphene which can be understood as electron-hole vortices with non-zero relative angular momentum of constituent particles. The condition of relative confinement depends on the electron-hole interaction strength, governed by the dielectric constant and the spread of the potential which is influenced by either the separation from the substrate or by screening. Therefore, our results might be useful for the ongoing search for the excitonic Mott transition in two-dimensional systems.\cite{Nikolaev08}

It is well-known that the zero-energy bound states are intimately related to resonant scattering of low-energy particles. This is reflected in the Levinson theorem,\cite{LevinsonTheorem} which is known to hold in the two-dimensional case.\cite{Portnoi97,Portnoi98} 
Our current work deals with zero-energy states only. Their mergence with the continuum when the potential is close to the critical strength and their influence on scattering and conductivity is a subject of future study, for which the knowledge of exact solutions for model potentials should provide a valuable tool.

\section*{Acknowledgments}
This work was supported by the EPSRC (CAD), the Millhayes foundation (DAS), the EU FP7 ITN Spinoptronics (Grant No. FP7-237252), FP7 IRSES projects SPINMET (Grant No. FP7-246784), TerACaN (Grant No. FP7-230778) and ROBOCON (Grant No. FP7-230832). We are grateful to R.~R.~Hartmann and N.~J.~Robinson for fruitful discussions, and we thank Prof. A.~J.~Makowski (Institute of Physics, Nicolaus Copernicus University, Toru{\'n}, Poland) for a critical reading of the manuscripts and valuable advice.

\end{document}